\documentclass[11pt]{article}
\RequirePackage{amsthm,amsmath,amsfonts,amssymb}
\RequirePackage[authoryear]{natbib}
\RequirePackage{color,graphicx}
\RequirePackage{graphicx}
\theoremstyle{definition}

\usepackage{caption}
\usepackage{subcaption}
\usepackage{bbm}
\usepackage{xcolor}
\usepackage{geometry}
\usepackage{natbib}
\usepackage[tableposition=top]{caption}
\setcitestyle{square, numbers,sort}
\usepackage[colorlinks = true,
            linkcolor = blue,
            urlcolor  = blue,
            citecolor = blue,
            anchorcolor = blue]{hyperref}
\usepackage{natbib}
\setcitestyle{authoryear,open={(},close={)}}
\providecommand{\keywords}[1]
{	\small
  \textbf{Keywords:} #1
}

\title{Mutually Exciting Point Processes \\ for Crowdfunding Platform Dynamics}

\author{Alexandra Djorno$^1$ and Forrest W. Crawford$^{1,2,3,4}$ \\[1em]
\normalsize
1. Department of Statistics \& Data Science, Yale University \\
2. Department of Biostatistics, Yale School of Public Health \\
3. Department of Ecology \& Evolutionary Biology, Yale University \\
4. Yale School of Management}

\date{}

\begin{document}
\maketitle

\begin{abstract}
Crowdfunding is a powerful tool for individuals or organizations seeking financial support from a vast audience. Despite widespread adoption, managers often lack information about dynamics of their platforms. Hawkes processes have been used to represent self-exciting behavior in a wide variety of empirical fields, but have not been applied to crowdfunding platforms in a way that could help managers understand the dynamics of users' engagement with the platform. In this paper, we extend the Hawkes process to capture important features of crowdfunding platform contributions and apply the model to analyze data from two donation-based platforms. For each user-item pair, the continuous-time conditional intensity is modeled as the superposition of a self-exciting baseline rate and a mutual excitation by preferential attachment, both depending on prior user engagement, and attenuated by a power law decay of user interest. The model is thus structured around two time-varying features -- contribution count and item popularity. We estimate parameters that govern the dynamics of contributions from 2,000 items and 164,000 users over several years. We identify a bottleneck in the user contribution pipeline, measure the force of item popularity, and characterize the decline in user interest over time. A contagion effect is introduced to assess the effect of item popularity on contribution rates. This mechanistic model lays the groundwork for enhanced crowdfunding platform monitoring based on evaluation of counterfactual scenarios and formulation of dynamics-aware recommendations. 
\end{abstract}
\keywords{Hawkes process, point process, mutually exciting process, contagion, crowdfunding}

\section{Introduction}

Over the past two decades, online crowdfunding platforms have emerged as an alternative to traditional financing from banks, markets, and venture capitalists \citep{schwienbacher2010crowdfunding}. They allow individuals and small businesses to raise funds from a large crowd of supporters through donations, rewards, loans or equity in time-limited campaigns \citep{cumming2020crowdfunding}. The generic term \emph{item} encompasses the diverse range of projects, products, films, art, music, entrepreneurial endeavors, real estate, and humanitarian initiatives undertaken by their \emph{owners} \citep{mollick2014dynamics}. A \emph{user} is defined broadly as any person engaging with the platform, either as a visitor or as a backer who makes a financial \emph{contribution}. Many studies have been conducted using data from Kickstarter \citep{kickstarterweb}, one of the largest reward-based crowdfunding platforms. Classification approaches have been employed by researchers to predict the success of an item, using both time series of users' contributions and static attributes of items \citep{etter2013launch, greenberg2013crowdfunding, chen2013kickpredict, rao2014emerging, chen2015will}. Survival models of items' success that accommodate censored observations have also been developed \citep{li2016project, jin2019estimating, dos2022reward}. Analyses of social network activity of owners and users have enriched models studying financial outcomes \citep{lu2014inferring, zvilichovsky2015playing}. The spreading mechanism of financial support has been explored through stochastic processes \citep{alaei2016dynamic, kindler2019early, ellman2022theory}. Additionally, crowdfunding platforms publicly display item descriptions, timing and amount of contributions, which can impact the decision-making of users \citep{burtch2013empirical, strausz2017theory, kim2018role, da2018beyond}. ``Rational herding'' has been observed among users, where items with higher funding levels and a larger number of users tend to attract more subsequent funding \citep{zhang2012rational, li2017catching, raab2020more, liu2021social}. This phenomenon aligns with theories of observational learning, social contagion and the wisdom of crowds \citep{banerjee1992simple, bikhchandani1992theory, aral2009distinguishing, mollick2016wisdom}. Intrinsic motivations, such as a sense of community with similar interest, can also drive user participation \citep{gerber2012crowdfunding, bagheri2019crowdfunding}. Familial and friendship connections often lead to early support of an item due to emotional ties with the owner \citep{lee2016financing, freedman2017information, kuppuswamy2018crowdfunding}. As their work often focuses on the success of individual items, researchers can provide advice and guidelines for item owners and managers, but do not necessarily examine their impact on the dynamics of the platform with a statistical model \citep{li2014dynamic, liu2015winner, kim2016all, rakesh2016probabilistic}. Moreover, these studies are often conducted on external or web-scraped data, without full information on user history and features, making it difficult to specify targeted interventions based on the number of prior contributions. 

Since its introduction in 1971, the Hawkes process \citep{hawkes1971spectra} has been used to represent temporal dependencies and cascading dynamics in a wide range of areas, such as earthquake aftershocks \citep{ogata1988statistical}, epidemics spread \citep{rizoiu2018sir}, social behaviors \citep{zhao2015seismic, kobayashi2016tideh, rizoiu2017hawkes, chen2018marked}, criminal activities \citep{mohler2011self}, and financial markets \citep{filimonov2012quantifying, bacry2015hawkes}. This self-exciting point process is useful for modeling situations where the occurrence of an arrival increases the likelihood of subsequent events. It can be viewed as a non-Markovian extension of the homogeneous Poisson process, with an arrival rate that depends on the history of events. Formally, a point process is a sequence of event times $\{t_1, t_2, \ldots \}$ corresponding to the counting process 
\[ N(t) = \sum_{k \geq 1} \boldsymbol{1}_{\{T_k \leq t \}} \]
for the number of occurrences up to time $t$. Features are attributes associated to each event time. Let $\mathcal{H}(t)$ be the history of events up to time $t$. A point process can be characterized by its continuous-time conditional intensity function 
\[ \lambda(t|\mathcal{H}(t)) = \lambda^*(t) = \lim_{h \to 0^+} \frac{\mathbb{E}[N(t+h) - N(t) | \mathcal{H}(t)]}{h}, \]
the expected rate of future events in $[t, t + \text{d}t]$, which depends only on past information \citep{daley2003introduction}. For the Hawkes process, the conditional intensity is given by 
\[ \lambda^*(t) = \lambda + \sum\limits_{t_k < t} \omega(t - t_k), \]
where $\lambda > 0$ is a baseline rate and $\omega :(0, \infty) \rightarrow [0, \infty] $ is an excitation function. Frequent choices for $\omega(\cdot)$ include exponential decay $\omega(t) = \alpha e^{-\delta t}$ and power law decay $\omega(t) = (t + \kappa )^{-(1 + \delta)}$, depending on the application \citep{laub2021elements}. The constant baseline rate can also be replaced by a time-varying background intensity \citep{chen2013inference}. In higher dimension, mutually exciting Hawkes processes are defined as a collection of $m$ counting processes $\{N_1(t), \ldots, N_m(t) \}$, where conditional intensities are influenced by events from other processes. They have been applied for modeling high frequency data \citep{bowsher2007modelling, bacry2013modelling},   financial contagion \citep{ait2015modeling}, viral diffusion \citep{yang2013mixture}, social influence \citep{zhou2013learning}, and temporal interactions on networks either at the node level \citep{fox2016modeling, passino2023graph} or for each edge \citep{blundell2012modelling, miscouridou2018modelling, huang2022mutually, passino2023mutually}. We are aware of only one paper that has applied Hawkes processes to crowdfunding data in order to differentiate between exogenous and endogenous contributions, but at the item level \citep{wang2021quantifying} and not for platform dynamics.
Working with internal data revealing returning users on different items would give the opportunity to define a Hawkes process for each user-item pair, and incorporate time-varying features dependent on user history for excitation phenomena. 

Our conversations with platform managers highlighted features associated with user engagement, item concentration, and user interest as key factors. Managers know how to select owners who will bring new users to their items. However, they have expressed concerns that most users contribute only once, while a very small fraction of ``cross-users'' donate to different items. Encouraging cross-backing is key to the long-term sustainability of platforms, as owners expect to receive contributions beyond their immediate network of family, friends, and acquaintances. These platforms charge a percentage fee of the total funds raised and compete with other crowdfunding platforms and online ``collection pot" sites that offer little visibility for an item but provide free or low-cost secure payment tools \citep{viotto2015competition}. We discovered that managers lack information about dynamics of contributions and effects of interventions on user engagement. This knowledge gap underscores the need for models that capture important features of user engagement and contribution to inform platform managers' strategic decision-making. 

In this paper, we study two crowdfunding platforms powered by the financial technology company \citet{mipiseweb} to uncover their contribution dynamics and quantify features that make users more likely to donate to items. We gathered 1,239 items with 77,000 users over nine years from Platform A, and 761 items with 87,000 users over eight years from Platform B. Drawing inspiration from epidemiological models \citep{crawford2020transmission, wu2018exposure}, we model contribution dynamics according to an extension of the Hawkes process that defines mutually exciting point processes for all user-item pairs. Key enhancements in our model include a baseline rate that is a function of the number of contributions already made by the user, a ``preferential attachment'' \citep{barabasi1999emergence} term that accounts for the relative popularity of the item, and a decreasing function representing the decay in the user interest in contributing. Given all these distinctive characteristics, our model is not a Hawkes process \emph{per se}, but defines a broader class of point processes that may be useful for modeling the rate of events between distinct sets of entities. We find the existence of a bottleneck between the first and second contributions for both platforms, with users likely to contribute to three or more items, once they have made two donations.  The relative popularity of an item has an important role on user's choice, with preferential attachment phenomenon. We introduce a contagion effect that decreases after a few contributions, slightly shifting user choices from following the crowd to donating to other items.  We estimate that user interest wanes rapidly over time, following a power law, characterized by fast decay with long-term memory, which can be explained by a behavior under financial constraints. This mechanistic model provides the methodological foundation for evaluation of counterfactual scenarios, which could help understand the impact of interventions of platform managers. Furthermore, the conditional intensity for each user-item pair paves the way for dynamics-aware recommendations that account for the platform evolution and are differentiated by individual features. 

\section{Data}
\label{sec: Data}

The data was obtained from \citet{mipiseweb}, a financial technology company that powers online crowdfunding platforms. We analyzed a dozen candidates and identified two platforms that had the necessary characteristics to support the goals of this study - significant volumes in donation along with a sufficiently high number of cross-users. Data from Platform A include 1,239 items, 77,000 users and 56,000 donations over nine years. Data from Platform B include 761 items, 87,000 users and 68,000 donations over eight years. For privacy, all the work was done with technical identifiers for items, users and contributions, which insured that all user and item data were anonymized. The variables for each item consist of item ID, type, start time, end time, while the variables for each user include user ID, registration time. Contributions of users to items are recorded with contribution ID, user ID, item ID, and contribution time.

We pre-processed the data by converting the date and time of each event into simple floating-point numbers, following the approach of \citet{laub2021elements}. Events such as user registration, item start, item end and contribution occur during discrete intervals, with many happening simultaneously. Because our model is based on continuous time, uniform noise is added at the millisecond level to uniquely identify each of these event. Leveraging the knowledge of user IDs, the number of items a user has contributed to is computed for each contribution. Additionally, the item relative popularity is obtained as the ratio of the number of users for that item to the number of users of all active items at time $t$. The elapsed time between the contribution and the user registration is also calculated. 

\begin{figure}
    \centering
    \includegraphics[width = 1\linewidth]{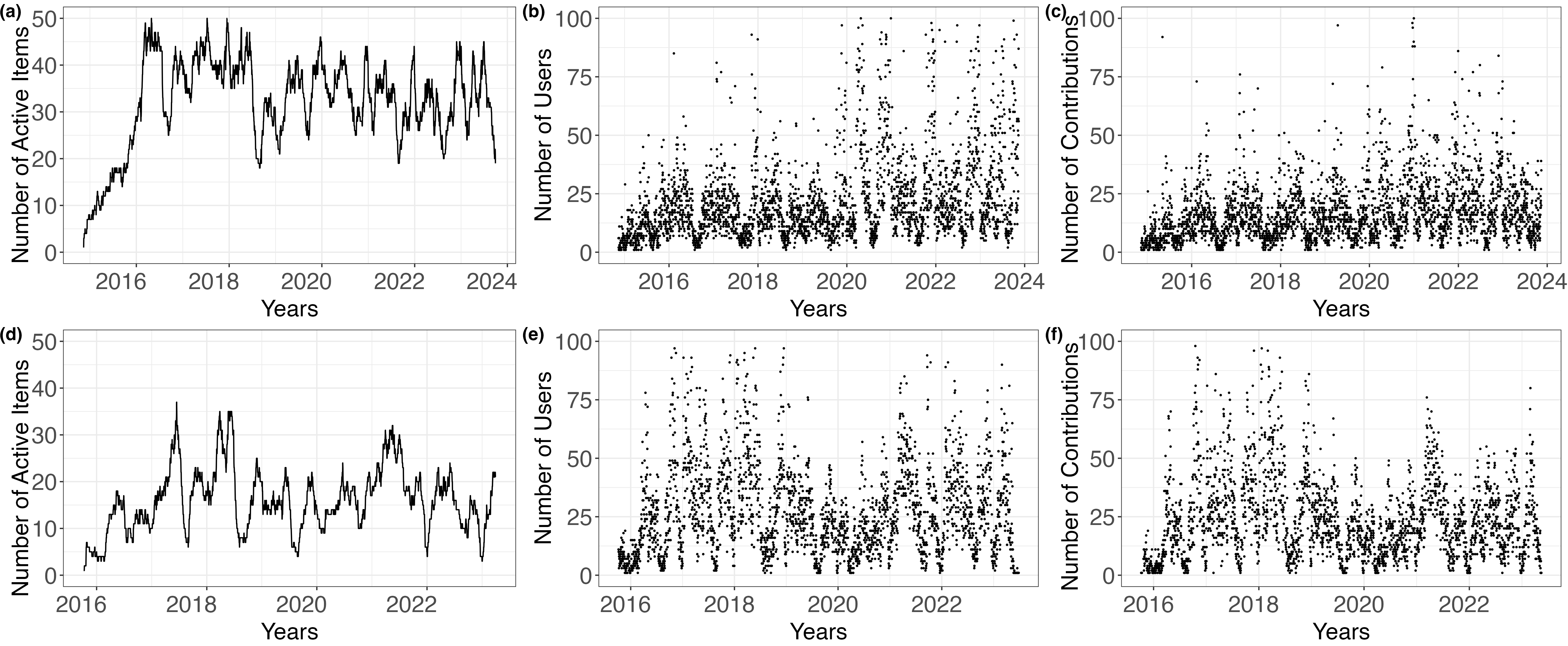}
    \caption{Dynamics of active items, users, and contributions per day on Platforms A and B.}
    \label{fig: Dynamics_platforms}
\end{figure}

\begin{figure}
    \centering
    \includegraphics[width = 1\linewidth]{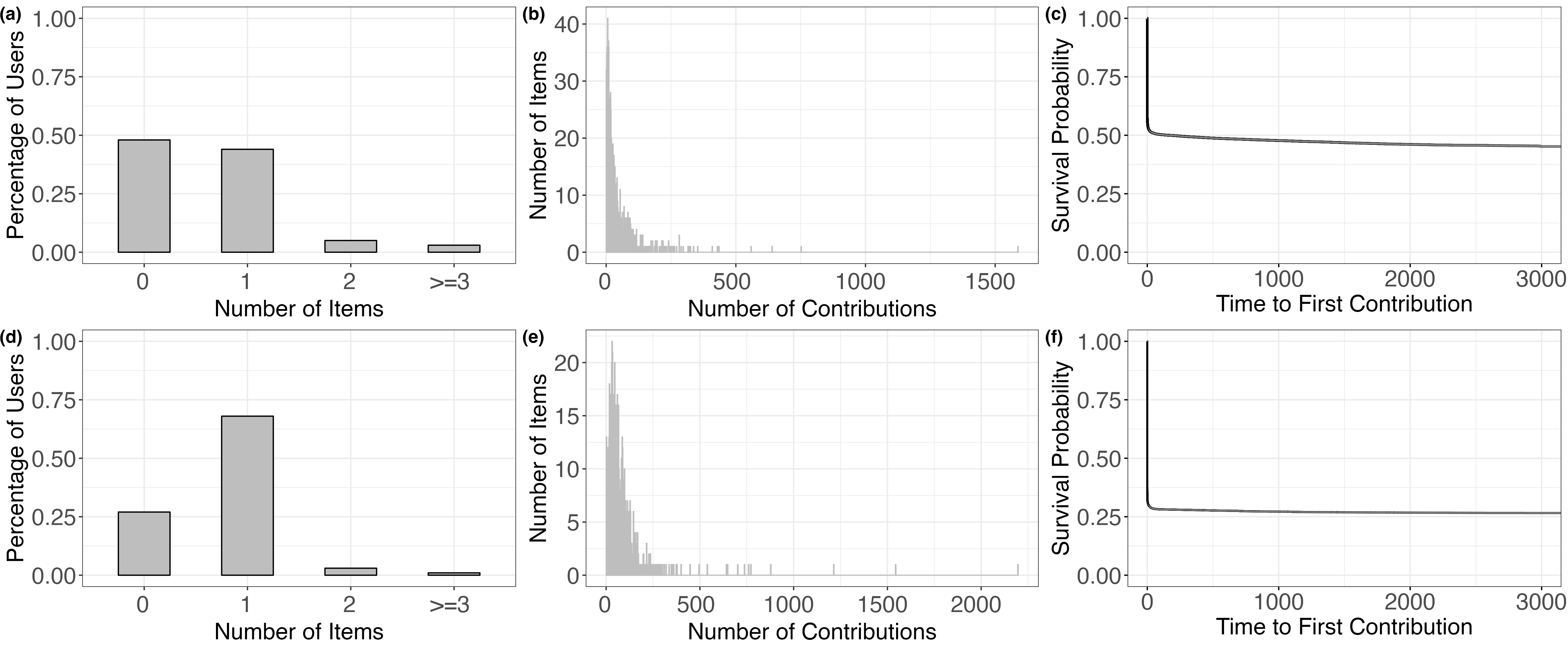}
    \caption{Distribution of users by the number of items, number of contributions per item, and the survival curve for the time to first contribution on Platforms A and B.}
    \label{fig: Frequencies}
\end{figure}

\begin{figure}
    \centering
    \includegraphics[width = 1\linewidth]{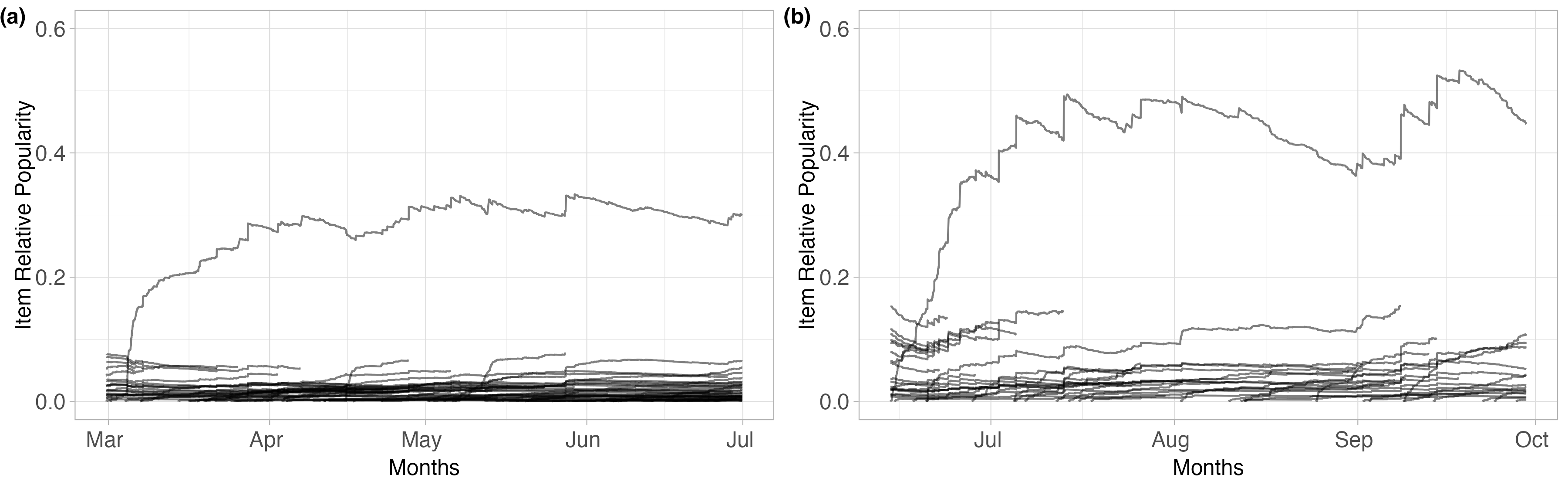}
    \caption{Item relative popularity on Platforms A and B, for each item, over time. Relative popularity is the ratio of the number of users who have contributed to the item to the number of users who have contributed to all active items at a given time.}
    \label{fig: Item_relative_popularity}
\end{figure}

Because items have a temporary duration on the platforms, with a median duration of 58 days on Platform A and 52 days on Platform B, Figures \ref{fig: Dynamics_platforms}a and \ref{fig: Dynamics_platforms}d show the number of active items at the same time. The tempo of new users per day is visualized in Figures \ref{fig: Dynamics_platforms}b and \ref{fig: Dynamics_platforms}e. Figures \ref{fig: Dynamics_platforms}c and \ref{fig: Dynamics_platforms}f highlight the contribution dynamics over time. The majority of users make zero or one contribution, while a small percentage of cross-users donate to multiple items as depicted in Figures \ref{fig: Frequencies}a and \ref{fig: Frequencies}d. Figures \ref{fig: Frequencies}b and \ref{fig: Frequencies}d show the number of contributions per item on both platforms, with half of items that attract 89\% and 82\% of the total number of contributions. Most the first contributions occur close to the user registration date, as shown in Figures \ref{fig: Frequencies}c and \ref{fig: Frequencies}f. Figures \ref{fig: Item_relative_popularity}a and \ref{fig: Item_relative_popularity}b illustrate the market share percentage of a highly popular item compared to the other active items at a given time.

\section{Methods}
\label{sec: Methods}

\subsection{Model Formulation}
\label{sec: Model Formulation}

Crowdfunding platforms dynamics consist not only of contribution flow, but also the influx of items and users. We construct a system of user-item interactions by considering four events: item start, item end, user registration, and user contribution to an item. At the platform launch time $t =0$, the sets of items $I(t)$ and users $U(t)$ are empty. A temporary item $i \in I(t)$ is considered ``active'' if $y_i \leq t \leq z_i$, where $y_i$ and $z_i$ are its start and end times. Items start with rate
\begin{equation}
    \zeta_{|I(t)|, |I(t)|+1}(t) = \phi, 
\end{equation}
and end with rate
\begin{equation}
        \zeta_{|I(t)|, |I(t)|-1}(t) = \mu |I(t)| 
\end{equation}
These crowdfunding platforms exhibit a distinct pattern, where new users register with the intention of supporting specific item owners. In other words, items drive the influx of users. The set of nodes $U(t)$ grows with a registration rate depending on the number of active items
\begin{equation}
\nu_{|U(t)|, |U(t)| + 1} (t) = \sigma|I(t)| 
\end{equation}
Our statistical model for contribution dynamics is based on the theory of point processes, to capture the dependence on past events and time-varying features. Contributions between each user-item pair are realizations of  $\mathbf{N} = \{N_{ui}(t)\}$, with a $|U(t)| \times |I(t)|$ time-varying matrix of conditional intensity functions $\boldsymbol{\lambda}^*(t) = \Big[\lambda_{ui}^*(t)\Big]$.  $N_{ui}(t)$ begins at $\max \{t_u^{(0)}, y_i \}$ when a pair $(u, i)$ comes into existence, and disappears when the temporary item ends at time $z_i$. Each entry $\lambda_{ui}^*(t) = \lambda_{ui}(t| \mathcal{H}_{ui}(t))$ consists of the superposition of a self-excitation function and a mutual excitation function, both attenuated by a power law decay of user interest. The contribution rate of user $u$ to item $i$ at time $t$ is 
\begin{equation}
\label{mod: reference model}
    \lambda_{ui}^*(t) = \left( \psi_{C_u(t)}+ \gamma_{C_u(t)} \frac{|U_i(t)|}{\sum\limits_{j \in I(t)} |U_j(t)|}  \right)\times \frac{1}{\left(t - t_u^{(0)} + \kappa\right)^{1 + \delta}}
\end{equation}
The conditional intensity function $\lambda_{ui}^*(t)$ is a combination of several elements. The first component is a self-excitation function. To track the progression through the various stages of the contribution pipeline, let $C_u(t) \in \{0, 1, 2, 3 \}$ be the contribution count, the number of contributions of user $u$ has made to distinct items by time $t$. Let $t_u^{(c)}$ be the time at which user $u$ makes their $c^{\text{th}}$ contribution, with $t_u^{(0)}$ representing the registration of the user on the platform. Let $\psi_{C_u(t)}$ be the baseline rate of user engagement as a function of the number of times $C_u(t)$ the user has contributed. 
\begin{equation} 
\psi_{C_u(t)} = \begin{cases} 
\psi_0 & t \in [t_u^{(0)}, t_u^{(1)}) \\
\psi_1 & t \in [t_u^{(1)}, t_u^{(2)}) \\
\psi_2 & t \in [t_u^{(2)}, t_u^{(3)}) \\
\psi_3 & t \geq t_u^{(3)}  
\end{cases}
\end{equation} 
Therefore, in contrast to the traditional framework, there is no exogenous component, as the baseline rate depends on the user engagement history. For the mutual excitation function, we take into account an important characteristic of crowdfunding platforms, which is the transparency of the number of contributions. The contribution intensity model permits a user's donation to be influenced by the relative popularity of an item
\[ \frac{|U_i(t)|}{\sum\limits_{j \in I(t)} |U_j(t)|},\]
defined as ratio of the number of users of item $i$ to the number of users of all active items at time $t$. This normalization prevents the number of users of an item from growing without bounds. We compute the relative cumulative number of users of an item instead of incorporating the observed sequence of past contribution times as in the Hawkes process, to avoid computational issues in the parameter estimation. Therefore, the item influence is measured with preferential attachment depending on contribution count $\gamma_{C_u(t)}$, to consider the impact of the crowd's decisions on user $u$ via the relative popularity of item $i$. 
\begin{equation} 
\gamma_{C_u(t)} = \begin{cases} 
\gamma_0 & t \in [t_u^{(0)}, t_u^{(1)}) \\
\gamma_1 & t \in [t_u^{(1)}, t_u^{(2)}) \\
\gamma_2 & t \in [t_u^{(2)}, t_u^{(3)}) \\
\gamma_3 & t \geq t_u^{(3)}  
\end{cases}
\end{equation} 
Additionally, the self and mutual excitation components are time-inhomogeneous, as they incorporate a power decay to materialize the waning user interest since his registration $t_u^{(0)}$. We found that a power law function provides a better fit than an exponential decay in capturing the fast decrease and long-term memory of the user interest. The exponent is equal to $1+\delta$ to guarantee that $\delta$ is positive, and the parameter $\kappa$ serves as a regularization constant to ensure the fraction is well-defined.

We introduce the contagion effect
\[ \beta_{C_u(t)} = \frac{\gamma_{C_u(t)}}{\psi_{C_u(t)}}\]
as the ratio of preferential attachment to user engagement, depending on the user contribution count. This metric offers a nuanced perspective on the evolution between user's and item's pulling forces. Monitoring the extent of concentration on larger items compared to smaller ones throughout the contribution pipeline could emerge as a key factor for the sustainability of a crowdfunding platform. If $\beta_{C_u(t)} > 1$, this indicates that the item influence, guided by the crowd, has more impact on the platform. Conversely, when $\beta_{C_u(t)}$ falls below 1, it signals a mind shift in users' decision-making to favor smaller items, which subsequently alters the dynamics on the platforms. 

\subsection{Statistical Estimation}
\label{sec: Statistical Estimation}

The integrated conditional intensity function is needed for the parameter estimation. Let \[\Lambda_{ui}(t) = \int_{0}^{t} \! \lambda_{ui}^*(v) \, dv\] be the compensator of a counting process $N_{ui}(t)$, interpreted as the conditional mean of $N_{ui}(t)$ given the past. The conditional intensity function $\lambda_{ui}^*(t)$ characterizes the random waiting time to contribution $T_{ui} \in (0, \infty]$, which begins when the user registers or when the item starts, whichever occurs later. To compute the compensator $\Lambda_{ui}(t)$ between user $u$ and item $i$ up to time $t$, let $\tau_{ui}^{(m)}$ denote the sequence of relevant times for the user-item pair, initialized with $\tau_{ui}^{(0)} = \max \{t_u^{(0)}, y_i \}$. The events of interest include the start and end times of the items, as well as all contributions by other users, during which the item relative popularity may vary. However, other users' registration times do not affect the likelihood of the waiting time $T_{ui}$.  Define $\eta_{ui}(t) = \max\limits_{m} \tau_{ui}^{(m)}$ such that $\tau_{ui}^{(m)} < t$. The integral needs to be partitioned into intervals of relevant times $[\tau_{ui}^{(m-1)}, \tau_{ui}^{(m)})$, due to the non-constant nature of the power law decay over $[0, t]$. While the sets $U(t), I(t), U_i(t)$ and $C_u(t)$ vary with time, they are right-continuous and count the number of events just before the time at which they are evaluated. This property allows the sets to be treated as constant, with only the power law decay term requiring integration over these intervals. Consequently, the compensator for user $u$ and item $i$ up to time $t$ is defined as follows:

\begin{equation}
\begin{split}
\label{eq: Compensator}
    \Lambda_{ui}(t) &= \sum\limits_{m: \tau_{ui}^{(m)} < t}
    \int_{\tau_{ui}^{(m-1)}}^{\tau_{ui}^{(m)}} \! \lambda_{ui}^*(v) \, dv + \int_{\tau_{ui}^{(\eta_{ui})}}^{t} \! \lambda_{ui}^*(v) \, dv  \\
    &= -\sum\limits_{m: \tau_{ui}^{(m)} < t} \left(\psi_{C_u\left(\tau_{ui}^{(m-1)}\right)} + \gamma_{C_u\left(\tau_{ui}^{(m-1)}\right)} \frac{\left|U_i\left(\tau_{ui}^{(m-1)}\right)\right|}{\sum\limits_{j \in I\left(\tau_{ui}^{(m-1)}\right)} \left|U_j\left(\tau_{ui}^{(m-1)}\right)\right|} \right) \times \\
    &\left(\frac{\left(\tau_{ui}^{(m)} - t_u^{(0)} + \kappa\right)^{-\delta}}{\delta} - \frac{\left(\tau_{ui}^{(m-1)} - t_u^{(0)} + \kappa\right)^{-\delta}}{\delta} \right)  \\
     &- \left(\psi_{C_u\left(\tau_{ui}^{(\eta_{ui})}\right)} +\gamma_{C_u\left(\tau_{ui}^{(\eta_{ui})}\right)} \frac{\left|U_i\left(\tau_{ui}^{(\eta_{ui})}\right)\right|}{\sum\limits_{j \in I\left(\tau_{ui}^{(\eta_{ui})}\right)} \left|U_j\left(\tau_{ui}^{(\eta_{ui})}\right)\right|} \right) \times \\
     &\left(\frac{\left(t - t_u^{(0)} + \kappa\right)^{-\delta}}{\delta} - \frac{\left(\tau_{ui}^{(\eta_{ui})} - t_u^{(0)} + \kappa \right)^{-\delta}}{\delta} \right) \\
\end{split}
\end{equation}
Proposition 7.2.III of \citet{daley2003introduction}
gives the general form of the likelihood function for univariate point processes as \[L = \left(\prod_{r =1}^{n} \lambda^*(s_r) \right) \exp \left(\int_{0}^{T} \! \lambda^*(v) \, dv \right)\] We can extend it to the setting of mutually exciting point processes in a system with start and end of items, along with registration of users. Specifically, let $s_1, \ldots, s_n$ be the event times observed continuously on the interval $[0, T]$, where $T = s_n$. Let $a_r = 1$ if event $r$ is the start of an item, $b_r = 1$ if $r$ is the end of an item, $d_r = 1$ if $r$ is the registration of a user, $k_{r, c} = 1$ if $r$ is a contribution, with $c \in \{0, 1, 2, 3\}$ to denote the count of contribution of user $u_r$ to item $i_r$. Let $a = \sum_{r = 1}^n a_r$ be the total number of new items, $b = \sum_{r = 1}^n b_r$ the total number of items that ended, $d = \sum_{r = 1}^n d_r$ the total number of new users, and $ k= \sum_{c=0}^{3} \sum_{r=1}^{n} k_{r, c}$ the total number of contributions. 

So, $a + b + d + k = n$, the total number of events in $[0, T]$. Let $u_r, r = 1, \ldots, n$ and $i_r, r = 1, \ldots, n$ denote the indices of users and items for event $r$. Let $\mathcal{U}$ and $\mathcal{I}$ be sets for running total of users and items, keeping track of all users and items who enter the platform during the interval $[0, T]$. Let $\mathcal{U}^{(c)}$, $c \in \{0, 1, 2, 3\}$, be the set of users who made a total of $c$ contributions in $[0, T]$, where $\mathcal{U}^{(3)}$ denotes the set of all users with three or more contributions. Let $\boldsymbol{\theta} = (\phi, \mu, \sigma, \psi_0, \psi_1, \psi_2, \psi_3, \gamma_0, \gamma_1, \gamma_2, \gamma_3, \kappa, \delta)$ be the vector of unknown parameters.  Then, the likelihood of the observed event times is
\begin{equation}
\begin{split}
    L(\boldsymbol{\theta}) &= 
    \prod_{r=1}^{n} \left(\phi^{a_r} \Big(\mu|I(s_r|\Big) ^{b_r} \Big(\sigma |I(s_r)| \Big)^{d_r} \prod_{c=0}^{3} \Big(\lambda_{u_r i_r}^*(s_r)\Big)^{k_{r, c}} \right) \\ 
    &\times \exp \left(- \phi s_n - \mu \sum\limits_{m: \tau^{(m)} \leq s_n} \Big|I \left(\tau^{(m-1)}\right)\Big| \left(\tau^{(m)} - \tau ^{(m-1)}\right) \right. \\
    &- \left. \sum\limits_{m: \tau^{(m)} \leq s_n} \sigma \Big|I\left(\tau^{(m-1)}\right)\Big| \left(\tau^{(m)} - \tau ^{(m-1)}\right) -\sum\limits_{u \in \mathcal{U}} \sum\limits_{i \in \mathcal{I}}\Lambda_{ui}(s_n)\right) 
\end{split}
\end{equation}
Appendix \ref{app: Maximum Likelihood Estimation} describes the log-likelihood, the derivations of partial derivatives and the Newton-Raphson algorithm to find the maximum likelihood estimates. 

\subsection{Simulations and Goodness-of-Fit Tests}
\label{sec: Simulations and Goodness-of-Fit Tests}

To evaluate the fit of the contribution model to the observed data, we conducted simulations using maximum likelihood estimates for the model. The approach consists of simulating the overall dynamics from $t=0$, by plugging in the estimated parameters to determine the shortest generated waiting time for all events of interest. The start and end of items are modeled using an exponential distribution with parameters $\phi$ and $\mu |I(t)|$. User registrations are also distributed according to an exponential distribution with parameter $\sigma |I(t)|$. The waiting time $x$ since the last event time $s$ for the mutually exciting point processes model of contributions can be obtained through inverse transform sampling (Algorithm 7.4.III of \citet{daley2003introduction}). Specifically, inverting the cumulative distribution function $F_s(x) = 1 - \exp \left(-\int_{s}^{s + x} \lambda^*(v)\, dv \ \right)$, defined in terms of the integrated conditional intensity, reduces to finding the root of $g(x) = F_s(x) - q$, where $q \sim$ Unif(0, 1). In particular, the waiting times until the next contribution of any user $u$ to any item $i$ is the root of the following function: \[ g_{ui}(x) = - \log(q) - \frac{1}{\delta}\left(\psi_{C_u(s)} + \gamma_{C_u(s)} \frac{\left|U_i(s)\right|}{\sum\limits_{j \in I(s)} |U_j(s)|}\right)\left[ \left(s+x - t_u^{(0)} +\kappa \right)^{-\delta} - \Bigl(s- t_u^{(0)} +\kappa \Bigr)^{-\delta}\right]\]
We simulated 100 replications the overall system dynamics over three years by varying the random seed and assessed the similarities with the observed cumulative trajectories.

Proposition 7.4.IV of \citet{daley2003introduction}, based on the random time change theorem \citep{brown2002time}, can be used to assess the goodness-of-fit of the mutually exciting point processes for the crowdfunding data. Specifically, consider an unbounded, increasing sequence of time points $\{t_1, t_2, \ldots, \}$ in the half-life $(0, \infty)$, and a monotonic, continuous compensator $\Lambda(\cdot)$ such that $\lim_{t\to\infty} \Lambda(t) = \infty$ almost surely. Then the transformed sequence $\{t_1^{*}, t_2^{*}, \ldots, \} = \{\Lambda(t_1), \Lambda(t_2), \ldots \}$ is a realization of a unit-rate Poisson process with probability one, if and only if the original sequence $\{t_r\}$ is a realization from the point process defined by $\Lambda(t)$. Thereafter, standard tests for homogeneous Poisson processes can be used on the transformed sequence $\{\Lambda(t_1), \Lambda(t_2), \ldots \}$. In particular, the interarrival times $\{t_1^{*}, t_2^{*} - t_1^{*}, t_3^{*} - t_2^{*}, \ldots, \}$ should follow an exponential distribution with rate 1. For $r = 2, \ldots, k$, where $k$ is the total number of contributions, the upper tail p-values 
\[p_r = \exp[-\Lambda(t_r) + \Lambda(t_{r-1})] = \exp\left(-\int_{t_{r-1}}^{t_r} \Lambda(v) \,dv\ \right)\]
should be uniformly distributed between 0 and 1. The Lewis test \citep{lewis1965some, kim2015power} is a more powerful statistical test, based on the ``conditional uniformity", which states that if $\{t_1^{*}, t_2^{*}, \ldots, t_k^{*}\}$ are arrival times for a unit rate Poisson process, then $\{t_1^{*}/t_k^{*}, t_2^{*}/t_k^{*}, \ldots, t_{k-1}^{*}/t_k^{*}\}$ are distributed as the order statistics of a sample from the uniform distribution between 0 and 1. Moreover, after checking the distribution, a qualitative autocorrelation test can be performed for independence of the interarrival times, by plotting the points $(V_{r}, V_{r-1})$, where $V_r = F(t_r^{*} - t_{r-1}^{*}) = 1 - \exp[-(t_r^{*} - t_{r-1}^{*})]$, as done in \citet{laub2021elements}. 

In the setting of mutually exciting point processes, we consider the overall compensator \[\Lambda(t) = \sum\limits_{u \in U(t)} \sum\limits_{i \in I(t)} \Lambda_{ui}(t)\] as the superposition of the waiting time processes across all $(u, i)$ pairs, and apply it to the realized contribution times. Specifically, all contributions arise from a single process, and can be transformed using the closed form expression in Equation \ref{eq: Compensator}. The function $\Lambda(t)$ considers all conditional intensity functions that are in effect at time $t$, even the unrealized ones, because of the competing risks perspective. Some pairs may have disappeared when an item ends, but their conditional intensity function is still in the cumulative sum, in the past.
Appendix \ref{app: Goodness-of-fit Tests} describes the bottom up dynamic programming approach to compute $\Lambda(t)$ at each realized contribution time in an efficient way.

\section{Results}
\label{sec: Results}

\subsection{Application to Two Crowdfunding Platforms}
\label{sec: Application to Two Crowdfunding Platforms}

This section presents results based on the estimated parameters of the model applied to two platforms A and B. We obtain a better understanding of crowdfunding dynamics, and observe similar patterns. Tables \ref{tab: Platform A} and \ref{tab: Platform B} show the estimated values of the parameters and the contagion effect along with their 95 \% confidence interval. The parameters $\phi$, $\mu$ and $\sigma$ control the number of active items and the influx of new users. The parameters $\psi_{C_u(t)}$, $\gamma_{C_u(t)}$ and $\beta_{C_u(t)}$  respectively correspond to the self-excitation, mutual excitation and the balance of the forces, for each value of the contribution count. $\delta$ indicates the pace at which the user interest declines.

\begin{table}
    \centering
\caption{Parameter Estimates and Contagion Effect for Platform A} 
\label{tab: Platform A}
\begin{tabular}{llll}
  \hline
 & Estimate & 2.5\% & 97.5\% \\ 
  \hline
$\phi$ & $3.77\times10^{-1}$ & $3.56\times10^{-1}$ & $3.98\times10^{-1}$ \\ 
  $\mu$ & $1.19\times10^{-2}$ & $1.12\times10^{-2}$ & $1.25\times10^{-2}$ \\ 
  $\sigma$ & $7.43\times10^{-1}$ & $7.38\times10^{-1}$ & $7.48\times10^{-1}$ \\ 
  $\psi_0$ & $9.28\times10^{-4}$ & $9.14\times10^{-4}$ & $9.43\times10^{-4}$ \\ 
  $\psi_1$ & $2.57\times10^{-4}$ & $2.47\times10^{-4}$ & $2.68\times10^{-4}$ \\ 
  $\psi_2$ & $7.80\times10^{-3}$ & $7.29\times10^{-3}$ & $8.30\times10^{-3}$ \\ 
  $\psi_3$ & $7.57\times10^{-2}$ & $7.30\times10^{-2}$ & $7.83\times10^{-2}$ \\ 
  $\gamma_0$ & $2.43\times10^{-2}$ & $2.38\times10^{-2}$ & $2.48\times10^{-2}$ \\ 
  $\gamma_1$ & $7.83\times10^{-3}$ & $7.48\times10^{-3}$ & $8.18\times10^{-3}$ \\ 
  $\gamma_2$ & $2.26\times10^{-1}$ & $2.10\times10^{-1}$ & $2.42\times10^{-1}$ \\ 
  $\gamma_3$ & $1.23\times10^{0}$ & $1.15\times10^{0}$ & $1.30\times10^{0}$ \\ 
  $\kappa$ & $1.18\times10^{-3}$ & $1.15\times10^{-3}$ & $1.20\times10^{-3}$ \\ 
  $\delta$ & $1.07\times10^{-1}$ & $1.05\times10^{-1}$ & $1.09\times10^{-1}$ \\ 
  $\beta_0$ & $2.62\times10^{1}$ & $2.57\times10^{1}$ & $2.67\times10^{1}$ \\ 
  $\beta_1$ & $3.04\times10^{1}$ & $2.91\times10^{1}$ & $3.18\times10^{1}$ \\ 
  $\beta_2$ & $2.90\times10^{1}$ & $2.69\times10^{1}$ & $3.11\times10^{1}$ \\ 
  $\beta_3$ & $1.62\times10^{1}$ & $1.52\times10^{1}$ & $1.72\times10^{1}$ \\ 
   \hline
\end{tabular}
\end{table}

\begin{table}
\centering
\caption{Parameter Estimates and Contagion Effect for Platform B} 
\label{tab: Platform B}
\begin{tabular}{llll}
  \hline
 & Estimate & 2.5\% & 97.5\% \\ 
  \hline
$\phi$ & $2.74\times10^{-1}$ & $2.54\times10^{-1}$ & $2.93\times10^{-1}$ \\ 
  $\mu$ & $1.73\times10^{-2}$ & $1.61\times10^{-2}$ & $1.86\times10^{-2}$ \\ 
  $\sigma$ & $1.92\times10^{0}$ & $1.91\times10^{0}$ & $1.94\times10^{0}$ \\ 
  $\psi_0$ & $2.20\times10^{-3}$ & $2.17\times10^{-3}$ & $2.23\times10^{-3}$ \\ 
  $\psi_1$ & $1.10\times10^{-4}$ & $1.03\times10^{-4}$ & $1.17\times10^{-4}$ \\ 
  $\psi_2$ & $9.07\times10^{-3}$ & $7.93\times10^{-3}$ & $1.02\times10^{-2}$ \\ 
  $\psi_3$ & $1.48\times10^{-1}$ & $1.36\times10^{-1}$ & $1.61\times10^{-1}$ \\ 
  $\gamma_0$ & $4.00\times10^{-2}$ & $3.95\times10^{-2}$ & $4.05\times10^{-2}$ \\ 
  $\gamma_1$ & $2.60\times10^{-3}$ & $2.47\times10^{-3}$ & $2.74\times10^{-3}$ \\ 
  $\gamma_2$ & $2.07\times10^{-1}$ & $1.86\times10^{-1}$ & $2.28\times10^{-1}$ \\ 
  $\gamma_3$ & $1.37\times10^{0}$ & $1.18\times10^{0}$ & $1.55\times10^{0}$ \\ 
  $\kappa$ & $2.27\times10^{-3}$ & $2.24\times10^{-3}$ & $2.30\times10^{-3}$ \\ 
  $\delta$ & $2.75\times10^{-1}$ & $2.73\times10^{-1}$ & $2.77\times10^{-1}$ \\ 
  $\beta_0$ & $1.82\times10^{1}$ & $1.79\times10^{1}$ & $1.84\times10^{1}$ \\ 
  $\beta_1$ & $2.37\times10^{1}$ & $2.25\times10^{1}$ & $2.49\times10^{1}$ \\ 
  $\beta_2$ & $2.28\times10^{1}$ & $2.05\times10^{1}$ & $2.52\times10^{1}$ \\ 
  $\beta_3$ & $9.21\times10^{0}$ & $7.95\times10^{0}$ & $1.05\times10^{1}$ \\ 
   \hline
\end{tabular}
\end{table}

The start rate for items $\phi$ is equal to 0.377 (95\% confidence interval=0.356, 0.398) and 0.274 (0.254, 0.293), respectively for Platform A and Platform B. On average, there are 137 (130, 145) and 100 (92.9, 107) new items per year, calculated as $\phi \times 365$. The end rate for items $\mu$ is estimated at 0.0119 (0.0112, 0.0125) and 0.0173 (0.0161, 0.0186). Consequently, the mean duration of an items, computed as $\frac{1}{\mu}$, is approximately 84.3 (79.9, 89.3) and 57.7 (53.9, 62.1) days. There are around 31.8 (31.8, 31.8) and 15.8 (15.8, 15.8) active items at a given time, determined by multiplying the start rate and the average lifetime of an item $\phi\times \frac{1}{\mu}$. The rate of new users $\sigma$ is equal to 0.743 (0.738, 0.748) and 1.92 (1.91, 1.94).

The individual examination of $\psi_{C_u(t)}$ and $\gamma_{C_u(t)}$ reveals how they are tied together in their evolution through the contribution pipeline. The user engagement parameters $\psi_{C_u(t)}$ represent baseline rates of contributions, with a breakdown by user count. These infinitesimal rates imply an additive change in the conditional intensity function, with units of contribution per user-item pair. For the first contribution $\psi_0$ is equal to 0.00092 (0.000914, 0.000943) and 0.00220 (0.00217, 0.00223), respectively for Platform A  for Platform B. For the second contribution, $\psi_1$ is estimated at $0.000257$ ($0.000247, 0.000268$) and $0.000110$ ($0.000103, 0.000117$). The conditional rate to the third contribution $\psi_2$ is found to be $0.00780$ ($0.00729, 0.00830$) and $0.00907$ ($0.00793, 0.0102$). Beyond the third contribution, $\psi_3$ is determined to be $0.0757$ ($0.0730, 0.0783$) and $0.148$ ($0.136, 0.161$). Concurrently, the parameters $\gamma_{C_u(t)}$ serve as metrics to quantify the influence of item popularity on user choice according to the number of prior contributions. They are multiplicative change per unit of relative popularity. $\gamma_0$ takes values of $0.0243$ ($0.0238, 0.0248$) and $0.0400$ (0.0395, $0.0405$). $\gamma_1$ is estimated to $0.00783$ ($0.00748, 0.00818$) and $0.00260$ ($0.00247$, $0.00274$). $\gamma_2$ is equal to $0.226$ ($0.210, 0.242$) and 0.207 $(0.186, 0.228$), while $\gamma_3$ is found to be $1.23$ ($1.15, 1.30$) and $1.37$ ($1.18, 1.55$). We discover a bottleneck both in the user engagement and the preferential attachment for the two platforms. There is a substantial drop in the values of $\psi_1$ and $\gamma_1$, which represent the conditional rates of transitioning from 1 to 2 items, when compared to $\psi_0$ and $\gamma_0$, suggesting the presence of a bottleneck in the contribution pipeline. In contrast, the values for $\psi_2$, $\gamma_2$ and $\psi_3$, $\gamma_3$ are significantly larger.  

The preceding parameters governing user choice must be contextualized with a waning user interest. By construction, the power law is characterized by a rapid decay but with a long-term memory effect. $\kappa$ is a regularization constant for the elapsed time between the contribution time and the user registration in units of days and is equal to $0.00118$ ($0.00115$, $0.00120$) and $0.00227$ ($0.00224, 0.00230$). Platform A has a lower $\delta$ value equal to $0.107$ ($0.105$, $0.109$) compared to $0.275$ ($0.273$, $0.277$) on Platform B, which indicates a higher user retention. This results in a 99.95\% decrease in user interest within one day of the registration, and a 99.99\% decline beyond four days for both platforms. 

The contribution conditional intensity for a user-item pair $\lambda_{ui}^*(t)$ is based on an interplay between the user propensity to contribute to several items (self-excitation) and the influence of others' contributions via the item relative popularity (mutual excitation). In this context, the contagion effect $\beta_{C_u(t)} = \frac{\gamma_{C_u(t)}}{\psi_{C_u(t)}}$ reflects this competition between preferential attachment on item $\gamma_{C_u(t)}$ and user engagement $\psi_{C_u(t)}$. $\beta_{C_u(t)} > 1$ suggests that the behavior of users is significantly influenced by the actions of the crowd. $\beta_{C_u(t)} < 1$ indicates that individual motivations play a stronger role in driving contributions. Tables \ref{tab: Platform A} and \ref{tab: Platform B} shows the evolution of the indicator according to the number of items a user has donated, with $C_u(t) \in \{0, 1, 2, 3\}$. The first three values are very strong, materializing the concentration around popular items. With the value of $\beta_0$, the conditional intensity from preferential attachment is found to be 26.2 (25.7, 26.7) and 18.2 (17.9, 18.4) times greater than the baseline rate for the first contribution on Platform A and Platform B, respectively. Similarly, $\beta_1$ exhibits values of 30.4 (29.1, 31.8) and 23.7 (22.5, 24.9), while $\beta_2$ is determined to be 29.0 (26.9, 31.1) and 22.8 (20.5, 25.2). There is a slight decline in the value of $\beta_3$, estimated at 16.2 (15.2, 17.2) and 9.21 (7.95, 10.5).   

\subsection{Simulation and Goodness-of Fit Results}
\label{sec: Simulation and Goodness-of Fit Results}

\begin{figure}
    \centering
    \includegraphics[width=1\linewidth]{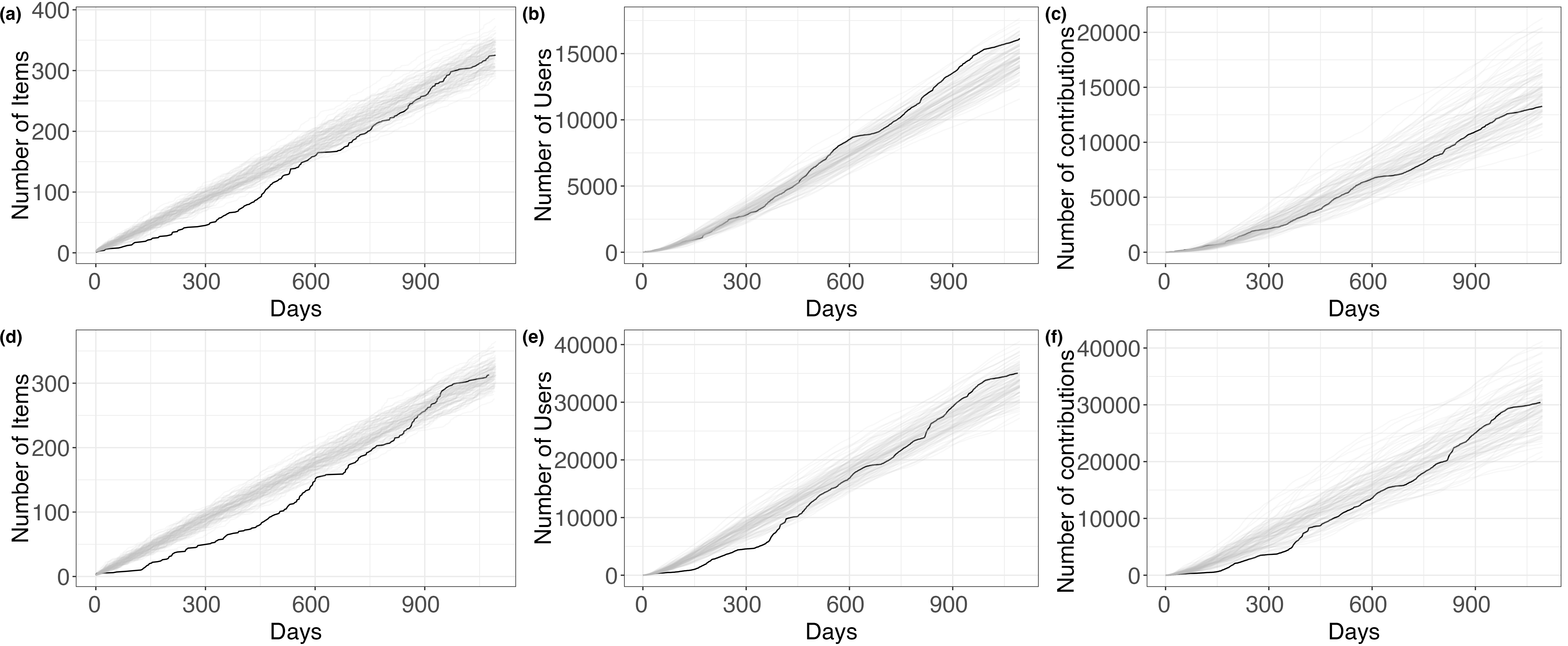}
    \caption{Results of full simulations of all platform dynamics, showing the observed number of items, users, and contributions overlaid on 100 simulated trajectories for Platforms A and B.}
    \label{fig: Simulations}
\end{figure}

Figure \ref{fig: Simulations} show that simulated and item, user and contribution frequencies (in grey) match closely the real data (in black) for Platforms A and B. 

\begin{figure}
    \centering
    \includegraphics[width=1\linewidth]{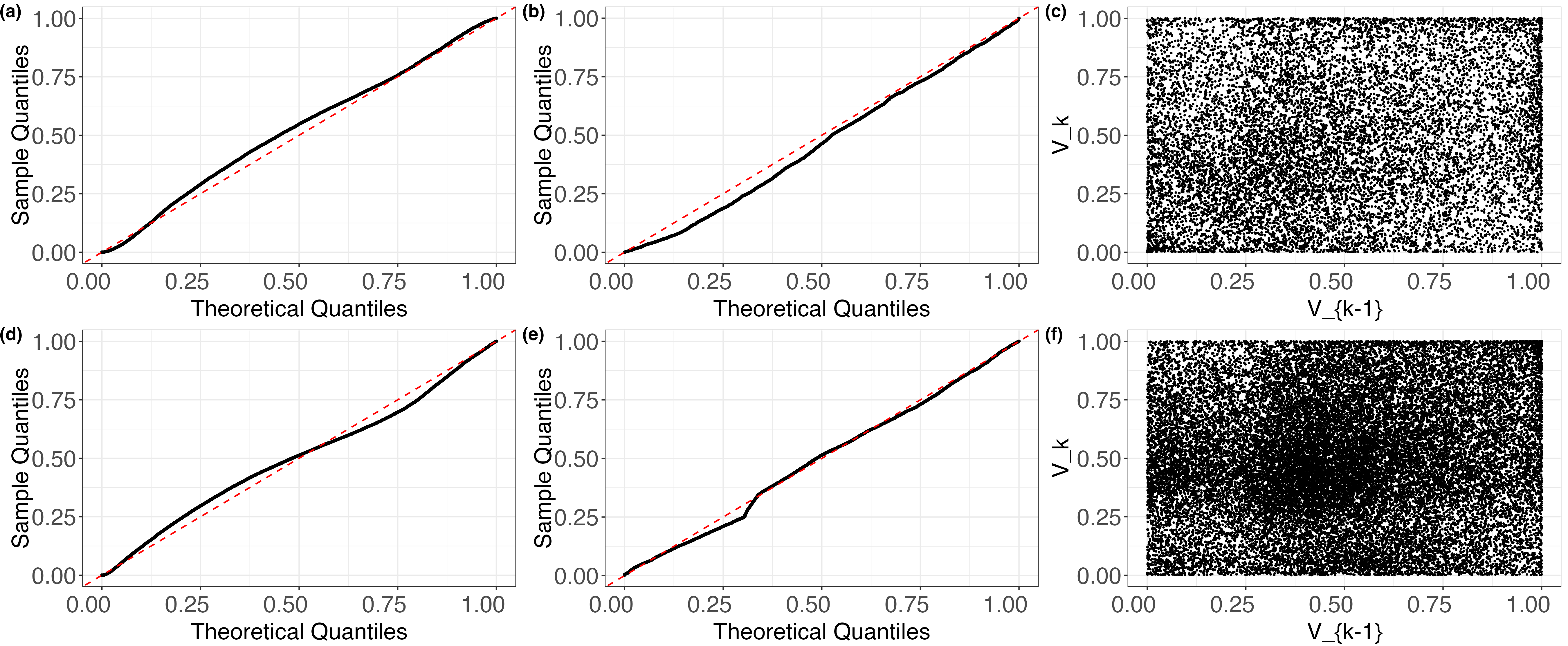}
    \caption{Q-Q plots to check the uniformity of the p-values and the transformed times conditioned on the last time, and autocorrelation plot of the interarrival times for Platforms A and B.}
    \label{fig: Goodness_of_fit_tests}
\end{figure}

Figures \ref{fig: Goodness_of_fit_tests}a and \ref{fig: Goodness_of_fit_tests}d present q-q plots to check whether the p-values described in Section \ref{sec: Simulations and Goodness-of-Fit Tests} are uniformly distributed between 0 and 1, for Platforms A and B, respectively. Figures \ref{fig: Goodness_of_fit_tests}b and \ref{fig: Goodness_of_fit_tests}e show q-q plots for the transformed conditional times vs a the theoretical quantiles of the uniform distribution. The distribution of the p-values and the conditional times is close to uniform. Finally, Figures \ref{fig: Goodness_of_fit_tests}c and \ref{fig: Goodness_of_fit_tests}d highlight a qualitative test for autocorrelation of the interarrival times. There does not seem to be any pattern, so the interarrival times are not autocorrelated. 

\subsection{Model Selection}
\label{sec: Model Selection}

In order to find the best model, we tested 13 different parametrizations on data from Platform A. This involved exploring various functional forms for the decay function, deriving multiple specifications for the relationship between independent contribution rates and preferential attachment contribution rates, and establishing some rates as functions of the number of prior contributions by a user. Table \ref{tab: Model selection} provides a comprehensive summary of the evaluated model functional forms, including the number of parameters involved, the variation of contribution count and the corresponding Akaike information criterion (AIC) \citep{akaike1974new} and Bayesian information criterion (BIC) \citep{schwarz1978estimating} values. The mutually exciting point processes model introduced in this paper emerges with the lowest AIC and BIC values among all models tested.

\begin{table}
\centering
\caption{Model Selection for Platform A} 
\begin{tabular}{lrlrr}
  \hline
Model & $\#$ params & $C_u(t)$ & AIC & BIC \\ 
  \hline
$(\psi_{C_u(t)} + \gamma_{C_u(t)}\frac{|U_i(t)|}{\sum\limits_{j \in I(t)} |U_j(t)|})(t - t_u + \kappa)^{-(1 + \delta)}$ & 10 & $0, 1, 2, 3$ & 438401 & 438516 \\ 
  $(\psi_{C_u(t)} + \gamma\frac{|U_i(t)|}{\sum\limits_{j \in I(t)} |U_j(t)|})(t - t_u + \kappa)^{-(1 + \delta)}$ & 7 & $0, 1, 2, 3$ & 451868 & 451957 \\ 
  $\psi_{C_u(t)}(t - t_u + \kappa)^{-(1 + \delta)}$ & 6 & $0, 1, 2, 3$ & 488067 & 488147 \\ 
  $(\psi_{C_u(t)} + \gamma\frac{|U_i(t)|}{\sum\limits_{j \in I(t)} |U_j(t)|})(t - t_u + \kappa)^{-(1 + \delta)}$ & 5 & $0, 1$ & 492514 & 492585 \\ 
  $\alpha + \beta (t - t_u + \kappa)^{-(1 + \delta)}$ & 4 &  & 493877 & 493940 \\ 
  $(\theta + \gamma\frac{|U_i(t)|}{\sum\limits_{j \in I(t)} |U_j(t)|})(t - t_u + \kappa)^{-(1 + \delta)}$ & 4 &  & 505019 & 505082 \\ 
  $\psi_{C_u(t)}(t - t_u + \kappa)^{-(1 + \delta)}$ & 4 & $0, 1$ & 514539 & 514601 \\ 
  $\gamma\frac{|U_i(t)|}{\sum\limits_{j \in I(t)} |U_j(t)|}(t - t_u + \kappa)^{-(1 + \delta)}$ & 3 &  & 526639 & 526692 \\ 
  $\theta(t - t_u + \kappa)^{-(1 + \delta)}$ & 3 &  & 536314 & 536368 \\ 
  $C_u(t) (\alpha + \beta e^{-\delta(t - t_u)} )$ & 3 & $\mathbb{N} $ & 547511 & 547565 \\ 
  $\theta C_u(t)(t - t_u + \kappa)^{-(1 + \delta)}$ & 3 & $\mathbb{N} $ & 549510 & 549563 \\ 
  $(\psi_{C_u(t)} + \gamma\frac{|U_i(t)|}{\sum\limits_{j \in I(t)} |U_j(t)|})e^{-\delta(t - t_u)}$ & 6 & $0, 1, 2, 3$ & 1023637 & 1023717 \\ 
  $(\theta + \gamma\frac{|U_i(t)|}{\sum\limits_{j \in I(t)} |U_j(t)|})e^{-\delta(t - t_u)}$ & 3 &  & 1097255 & 1097309 \\ 
   \hline
\end{tabular}
\label{tab: Model selection}
\end{table}

\section{Discussion}
\label{sec: Discussion}

We propose mutually exciting point processes, included in a system with temporary items and an influx of new users, to understand crowdfunding platform dynamics. Our user-centric approach differs from existing studies that predominantly adopt an item-centric focus to predict fundraising success. Even the preferential attachment is considered through the user's perspective with contribution count $C_u(t)$. By combining evolving system of items and users, self and mutual excitation, long-term memory, along with time-varying features (contribution count, item popularity), this generative continuous-time stochastic model allows estimating specific parameter values that can guide platform manager actions. 

Through the findings, we can advise managers seeking to optimize and improve their operational performance. We find a bottleneck both in the user engagement and the preferential attachment for the two platforms. As a consequence, users rarely contribute more than once, but if they donate to 2 items, referred to as cross-users, they are likely to contribute to 3 items or more. For managers with constrained resources, a suggestion could be to stimulate one-item users through effective communication or incentives, and redirect efforts away from cross-users who have already established a contributing habit. This trade-off becomes a key factor for marketing investments. The first three values of the contagion effect are very high, characterizing the concentration around popular items. This reinforcing loop leads to a ``rich-get-richer" phenomenon, which results in a scale-free network, with item hubs that attract a substantial number of users from the platforms. Managers can implement specific strategies to leverage this momentum, such as continuously attracting potential popular items to capitalize on the bandwagon effect, assuming that this popularity is an indicator for quality or potential success. However, this approach should be exercised judiciously to avoid potential drawbacks arising from concentration. The drop in the value of $\beta_{3}$ can be interpreted as a slight mind shift from a choice driven by mutual excitation following the crowd, towards a self-exciting individual motivation to donate to other presented items, regardless of the size. From a marketing perspective, this kind of spillover mechanism from popular items could serve as a compelling argument to attract owners of smaller items, allowing them to harness dynamics beyond their personal network. The user interest declines extremely rapidly, with a long-term memory. This behavior is similar to the waning interest observed in microblogs and social media, \citep{gao2015modeling, yang2019non}, and is typical of random human activity, which has been shown to have a bursting nature, characterized by a heavy-tailed distribution \citep{barabasi2005origin, newman2005power}. Therefore, if managers aim to make impactful recommendations, the timing is crucial and should align closely with user registration. In simpler terms, it is better to strike while the iron is hot, to capture and maintain user interest. 

Our approach and data are subject to limitations. This modeling was only possible thanks to internal data availability, in particular, the contribution count $C_u(t)$ based on user histories. Moreover, the complexity of this model, with its 10 parameters, may lead to difficulties in parameter estimation for relatively small datasets. Several  platforms were analyzed during this study, and only two had the minimum number of cross-users necessary for convergence to the maximum likelihood estimates. As all possible user-item pairs over several years have to be considered in the parameter estimation, our algorithm needs to process over 25 million rows. Scaling the framework for much bigger datasets will require substantial computational and memory resources. In addition, parametric modeling assumptions, such as constant self-excitation and mutual excitation dependent on the contribution count along with conditional independence of waiting times to contribution, were introduced to guarantee that the parameters are both interpretable and identifiable. Inferences could be biased if the conditional intensity function is incorrectly specified. If the level of returning users is increased significantly, we could also adapt the power law decay to account for the elapsed time since the last user contribution instead of user registration. 

With the dynamics now modeled in a mechanistic framework, we can extend our focus to broader aspects of platform health for long-term sustainability, through enhanced monitoring. The natural progression leads us to consider the design of a dynamics-aware recommender system. By leveraging the conditional intensity function, we have an opportunity to transcend traditional methods rooted in content and collaborative filtering \citep{ricci2010introduction}. The versatility of the modeling allows considering additional individual features, such as item category or geographical distance between item owner and user, for more personalized recommendations. In order to identify powerful levers, we need to delve into the evaluation of counterfactual scenarios involving nonlinear effects. For instance, within a bottleneck context, what would happen if the rate of second contribution is increased by $x\%$? To refine interventions, it is necessary to establish thresholds. For instance, at which level would the contagion effect be too high, prompting to strategically boost smaller items to limit concentration and maintain a balanced system? Equipped with insights derived from a statistical model, platform managers will possess a toolkit to guide interventions, supplementing their intuition and experience. This work not only advances our understanding of platform dynamics with point processes, but also underscores the reciprocal learning opportunities that exist between academia and business. 

\textbf{Acknowledgements}

We are grateful to Michel Ivanovsky, Eric Didio and Timothée Tixier for providing the data and for insightful conversations about the crowdfunding business model. We thank Andrew Barron, Sekhar Tatikonda, and Joe Chang for helpful comments on the manuscript. 

\bibstyle{plain}
\bibliography{paperbibliography}

\section*{Appendix}
\begin{appendix}
\section{Maximum Likelihood Estimation}
\label{app: Maximum Likelihood Estimation}

Substituting the definitions of $\lambda_{ui}^*(t)$ and $\Lambda_{ui}(t)$, the log-likelihood becomes:
\begin{equation*}
    \begin{split}
        &\ell(\boldsymbol{\theta}) = \sum_{r=1}^{n} \Bigg(a_r \log(\phi) + b_r \log(\mu |I(s_r)|) + d_r \log(\sigma \left|I(s_r) \right|) \\
        &+ \sum_{c=1}^{3} k_{r,c}\log\Bigl[\Bigl(\psi_{C_{u_r}(s_r)} + \gamma_{C_{u_r}(s_r)} \frac{|U_{i_r}(s_r)|}{\sum\limits_{j \in I(s_r)} |U_j(s_r)|} \Bigr) \frac{1}{(s_r - t_{u_r}^{(0)} + \kappa)^{1 + \delta}} \Bigr] \Bigg)\\
        &- \phi T - \mu \sum\limits_{m: \tau^{(m)} \leq T} \left|I(\tau^{(m-1)})\right| \left(\tau^{(m)} - \tau ^{(m-1)}\right) -  \sigma \sum\limits_{m: \tau^{(m)} \leq T} \left|I(\tau^{(m-1)}) \right| \left(\tau^{(m)} - \tau ^{(m-1)}\right)  \\
        &+ \sum\limits_{u \in \mathcal{U}} \sum\limits_{i \in \mathcal{I}} \sum_{\substack{m: \tau_{ui}^{(m)} < \min\{z_i, T\} \\ m: \tau_{ui}^{(m)} \geq \max\{t_u^{(0)}, y_i\}}} \left(\psi_{C_u\left(\tau_{ui}^{(m-1)}\right)} + \gamma_{C_u\left(\tau_{ui}^{(m-1)}\right)} \frac{\left|U_i\left(\tau_{ui}^{(m-1)}\right)\right|}{\sum\limits_{j \in I\left(\tau_{ui}^{(m-1)}\right)} \left|U_j\left(\tau_{ui}^{(m-1)}\right)\right|} \right) \times \\
        &\left(\frac{\left(\tau_{ui}^{(m)} - t_u^{(0)} + \kappa\right)^{-\delta}}{\delta} - \frac{\left(\tau_{ui}^{(m-1)} - t_u^{(0)} + \kappa\right)^{-\delta}}{\delta} \right)  \\
        &+ \left(\psi_{C_u\left(\tau_{ui}^{(\eta_{ui})}\right)}  + \gamma_{C_u\left(\tau_{ui}^{(\eta_{ui})}\right)} \frac{\left|U_i\left(\tau_{ui}^{(\eta_{ui})}\right)\right|}{\sum\limits_{j \in I\left(\tau_{ui}^{(\eta_{ui})}\right)} \left|U_j\left(\tau_{ui}^{(\eta_{ui})}\right)\right|} \right) \times \left(\frac{\left(T - t_u^{(0)} + \kappa\right)^{-\delta}}{\delta} - \frac{\left(\tau_{ui}^{(\eta_{ui})} - t_u^{(0)} + \kappa\right)^{-\delta}}{\delta} \right)
    \end{split}
\end{equation*}
For each platform, we estimate the vector of unknown parameters $\boldsymbol{\theta}$ by maximum likelihood. Our goal is to obtain the estimate $\boldsymbol{\hat\theta} = \underset{\boldsymbol{\theta}}{\operatorname{argmax}} \ \ell(\boldsymbol{\theta})$. However, the closed-formed expressions for the parameters governing contribution the conditional intensity are not available. Consequently, we use the Newton-Raphson method, reducing maximization of the log-likelihood to root-finding for $\nabla_{\boldsymbol{\theta}} \ell(\tilde{\boldsymbol{\theta}})$. Specifically, the update for the Newton-Raphson algorithm is given by 
\[\tilde{\boldsymbol{\theta}}^{(m+1)} = \tilde{\boldsymbol{\theta}}^{(m)} - \left[\nabla{^2}_{\boldsymbol{\theta}} \ell\left(\tilde{\boldsymbol{\theta}}^{(m)}\right)\right]^{-1} \nabla_{\boldsymbol{\theta}} \ell\left(\tilde{\boldsymbol{\theta}}^{(m)}\right)\]
where $\tilde{\boldsymbol{\theta}}^{(m)}$ is the $m^{\text{th}}$ iterate. Numerical convergence is achieved at iteration $m$ when $\nabla_{\boldsymbol{\theta}} \ell\left(\tilde{\boldsymbol{\theta}}^{(m)}\right) \approx \boldsymbol{0}$ and $ || \tilde{\boldsymbol{\theta}}^{(m)} - \tilde{\boldsymbol{\theta}}^{(m-1)}|| \approx 0$. The maximum likelihood estimate is then $\boldsymbol{\hat{\theta}} = \tilde{\boldsymbol{\theta}}^{(m)}$. Using the asymptotic normality of the maximum likelihood estimates, standard errors are given by $\text{SE}(\boldsymbol{\hat{\theta}}) = \sqrt{-\text{diag} \left[\nabla{^2}_{\boldsymbol{\theta}} \ell\left(\tilde{\boldsymbol{\theta}}^{(m)}\right)\right]^{-1}}$, 
where diag$(\cdot)$ is the diagonal of its matrix argument. 

We compute derivatives of $\ell$ with respect to the elements of $\boldsymbol{\theta}$. There are closed-form expression for the parameters governing the item flow $\phi$ and $\mu$, and the user registration $\sigma$.
\begin{equation*}
\begin{split}
    &\frac{\partial{\ell}}{\partial{\phi}} = \frac{\sum_{r=1}^{n} a_r}{\phi} - T ; \frac{\partial{\ell}}{\partial{\mu}}= \frac{\sum_{r=1}^{n} b_j}{\mu} - \int_0^{t} \! |I(v)|\, \mathrm{d}v; \frac{\partial{\ell}}{\partial{\sigma}} = \frac{\sum_{r=1}^{n} d_r}{\sigma} - \int_0^{t_n} \! |I(v)|\, \mathrm{d}v \\
    & \Rightarrow \hat{\phi} = \frac{\sum_{r=1}^{n} a_r}{t} ; \ \hat{\mu} = \frac{\sum_{r=1}^{n} b_r}{\int_0^{t} \! |I(v)| \, \mathrm{d}v} ; \ \hat{\sigma} = \frac{\sum_{r=1}^{n} d_r}{\int_0^{t} \! |I(v)|\, \mathrm{d}v}
\end{split}
\end{equation*}
Let $t_u^{(c)}$ be the time of the $c^{\text{th}}$ contribution of user $u$.
Then, for all $c \in \{ 0, 1, 2, 3 \}$:
\begin{equation*}
\begin{split}
    \frac{\partial{\ell}}{\partial{\psi_{c}}} &= \sum_{r = 1}^n \frac{k_{r, c}}{\Bigl(\psi_{c} + \gamma_{c} \frac{|U_{i_r}(s_r)|}{\sum\limits_{j \in I(s_r)} |U_j(r_j)|} \Bigr)} + \sum\limits_{u \in \mathcal{U}^{(c)}} \sum\limits_{i \in \mathcal{I}} \frac{1}{\delta} \times \\
    &\Bigg[\Bigl(\min \{t_u^{(c+1)} \boldsymbol{1}_{\{ c < 3\}}, T\boldsymbol{1}_{\{ c = 3\}}, z_i\} - t_u^{(0)} + \kappa \Bigr)^{-\delta} - \Bigl(\max \{t_u^{(c)}, y_i \} - t_u^{(0)} + \kappa\Bigl)^{-\delta}\Bigg]  \\
\end{split}
\end{equation*}
\begin{equation*}
    \begin{split}
         \frac{\partial{\ell}}{\partial{\gamma_c}} & = \sum_{r = 1}^n \frac{k_{r, c} \frac{|U_{i_r}(s_r)|}{\sum\limits_{j \in I(s_r)} |U_j(s_r)|} }{\Bigl(\psi_c + \gamma_c \frac{|U_{i_r}(s_r)|}{\sum\limits_{j \in I(s_r)} |U_j(s_r)|} \Bigr)} \\
        &+ \frac{1}{\delta}\sum\limits_{u \in \mathcal{U}^{(c)}} \Bigg[\Bigl(\min \{t_u^{(c+1)}\boldsymbol{1}_{\{ c < 3\}}, T\boldsymbol{1}_{\{ c = 3\}}\} - t_u^{(0)} + \kappa \Bigr)^{-\delta} - \Bigl(t_u^{(c)} - t_u^{(0)} + \kappa \Bigr)^{-\delta}\Bigg]
    \end{split}
\end{equation*}
\begin{equation*}
\begin{split}
    &\frac{\partial{\ell}}{\partial{\kappa}}= -\sum_{c = 0}^{3}\sum_{r = 1}^n \frac{k_{r, c}(1 + \delta)}{\Bigl(s_r - t_{u_r}^{(0)} + \kappa \Bigr)} - \sum_{c=0}^{3} \sum\limits_{u \in \mathcal{U}^{(c)}} \sum\limits_{i \in \mathcal{I}} \psi_{c} \times \\
    &\Bigg[\Bigl(\min \{t_u^{(c+1)}\boldsymbol{1}_{\{ c < 3\}}, T\boldsymbol{1}_{\{ c = 3\}}, z_i \} - t_u^{(0)} + \kappa\Bigr)^{-(1 + \delta)}- (\max \{t_u^{(c)}, y_i \} - t_u^{(0)} + \kappa)^{-(1 +\delta)}\Bigg] \\
    &- \sum_{c=0}^{3}\sum\limits_{u \in \mathcal{U}^{(c)}} \gamma_c \Bigg[ \Bigl(\min \{t_u^{(c+1)}\boldsymbol{1}_{\{ c < 3\}}, T\boldsymbol{1}_{\{ c = 3\}}\} - t_u^{(0)} + \kappa\Bigr)^{-(1 + \delta)} - \Bigl(t_u^{(c)} - t_u^{(0)} + \kappa\Bigr)^{-(1 +\delta)}\Bigg]
\end{split}
\end{equation*}
\begin{equation*}
    \begin{split}
    &\frac{\partial{\ell}}{\partial{\delta}}= -\sum_{c=0}^{3} \sum_{r = 1}^n k_{r, c} \ln(s_r - t_{u_r}^{(0)} + \kappa) \\
    &-\sum_{c=0}^{3}\sum\limits_{u \in \mathcal{U}^{(c)}} \sum\limits_{i \in \mathcal{I}} \frac{\psi_c}{\delta^2} \Bigg[\Bigl(\min \{t_u^{(c+1)}\boldsymbol{1}_{\{ c < 3\}}, T\boldsymbol{1}_{\{ c = 3\}}, z_i \} - t_u^{(0)} + \kappa \Bigr)^{-\delta} \times \\
    &\Bigl[\delta \ln \Bigl(\min \{t_u^{(c+1)}\boldsymbol{1}_{\{ c < 3\}}, T\boldsymbol{1}_{\{ c = 3\}}, z_i \} - t_u^{(0)} + \kappa \Bigr)+1 \Bigr]\\
    &- \Bigl(\max \{t_u^{(c)}, y_i \} - t_u^{(0)} + \kappa \Bigr)^{-\delta} \Big[\delta \ln \Bigl(\max \{t_u^{(c)}, y_i \} - t_u^{(0)} + \kappa \Bigr)  + 1 \Big] \Bigg]  \\
    &-\sum_{c=0}^{3}\sum\limits_{u \in \mathcal{U}^{(c)}} \frac{\gamma_c}{\delta^2} \Bigg[ \Bigl(\min \{t_u^{(c+1)} \boldsymbol{1}_{\{ c < 3\}}, T\boldsymbol{1}_{\{ c = 3\}}\} - t_u^{(0)} + \kappa \Bigr)^{-\delta} \times \\
    &\Big[\delta \ln \Bigl(\min \{t_u^{(c+1)}\boldsymbol{1}_{\{ c < 3\}},T\boldsymbol{1}_{\{ c = 3\}}\} - t_u^{(0)} + \kappa \Bigr)+1 \Big]\\
    &- \Bigl(t_u^{(c)} - t_u^{(0)} + \kappa \Bigr)^{-\delta} \Bigr[\delta \ln \Bigl(t_u^{(c)} - t_u^{(0)} + \kappa \Bigr) + 1\Bigl] \Bigg]  \\
    \end{split}
\end{equation*}
\begin{equation*}
    \frac{\partial^2{\ell}}{\partial{\psi_c \psi_{c'}}} = 
    \begin{cases}
    -\sum_{r = 1}^{n} \dfrac{k_{r, c}^2}{\Bigl(\psi_c + \gamma_c \frac{|U_{i_r}(s_r)|}{\sum\limits_{j \in I(s_r)} |U_j(s_r)|} \Bigr)^2} & \text{if $c = c'$} \\
    0 & \text{if $c \neq c'$} 
    \end{cases}
\end{equation*}
\begin{equation*}
    \frac{\partial^2{\ell}}{\partial{\gamma_c \gamma_{c'}}} = 
    \begin{cases}
    -\sum_{r = 1}^n \dfrac{k_{r, c} \Bigl(\frac{|U_{i_r}(s_r)|}{\sum\limits_{j \in I_{u_r}(s_r)} |U_j(s_r)|} \Bigr)^2}{\Bigl(\psi_c + \gamma_c \frac{|U_{i_r}(s_r)|}{\sum\limits_{j \in I(s_r)} |U_j(s_r)|} \Bigr)^2} & \text{if $c = c'$} \\
    0 & \text{if $c \neq c'$}
    \end{cases}
\end{equation*}
\begin{equation*}
    \begin{split}
     &\frac{\partial^2{\ell}}{\partial{\kappa^2}} = \sum_{c=0}^{3}\sum_{r = 1}^n \frac{k_{r, c} (1 + \delta)}{\Bigl(s_r - t_{u_r}^{(0)} + \kappa \Bigr)^2} + \sum_{c=0}^{3} \sum\limits_{u \in \mathcal{U}^{(c)}} \sum\limits_{i \in \mathcal{I}}  \psi_c \Bigl(1+\delta \Bigr) \times \\
     &\Bigg[\Bigl(\min \{t_u^{(c+1)}\boldsymbol{1}_{\{ c < 3\}}, T\boldsymbol{1}_{\{ c = 3\}}, z_i\} - t_u^{(0)} + \kappa)^{-(2 + \delta)} \Bigr) - \Bigl(\max \{t_u^{(c)}, y_i\} - t_u^{(0)} + \kappa\Bigr)^{-(2 +\delta)} \Bigg] \\
     &+ \sum_{c=0}^{3} \sum\limits_{u \in \mathcal{U}^{(c)}}\gamma_c \Bigl(1+\delta \Bigr) \Bigg[ \Bigl(\min \{t_u^{(c+1)}\boldsymbol{1}_{\{ c < 3\}}, T\boldsymbol{1}_{\{ c = 3\}}\} - t_u^{(0)} + \kappa\Bigr)^{-(2 + \delta)} - \Bigl(t_u^{(c)} - t_u^{(0)} + \kappa\Bigr)^{-(2 +\delta)}\Bigg]\\
    \end{split}
\end{equation*}
\begin{equation*}
    \begin{split}
     &\frac{\partial^2{\ell}}{\partial{\delta^2}} = \sum_{c=0}^{3} \sum\limits_{u \in \mathcal{U}^{(c)}} \sum\limits_{i \in \mathcal{I}} \frac{\psi_c}{\delta^3} \times \Bigg[\Bigl(\min \{t_u^{(c+1)}\boldsymbol{1}_{\{ c < 3\}}, T\boldsymbol{1}_{\{ c = 3\}}, z_i \} - t_u^{(0)} + \kappa \Bigr)^{-\delta} \times \\
     & \Bigg(\Bigl[\delta \ln \Bigl(\min \{t_u^{(c+1)}\boldsymbol{1}_{\{ c < 3\}}, T\boldsymbol{1}_{\{ c = 3\}}, z_i \} - t_u^{(0)} + \kappa \Bigr) + 1\Bigr]^2 + 1 \Bigg) \\
     &-\Bigl(\max \{t_u^{(c)}, y_i \} - t_u^{(0)} + \kappa \Bigr)^{-\delta} \Bigg(\Bigl[\delta \ln \Bigl(\max \{t_u^{(c)}, y_i \} - t_u^{(0)} + \kappa \Bigr) + 1 \Bigr]^2 + 1 \Bigg) \Bigg]\\
    &-\sum_{c=0}^{3}\sum\limits_{u \in \mathcal{U}^{(c)}} \frac{\gamma_c}{\delta^3} \Bigg[ \Bigl(\min \{t_u^{(c+1)} \boldsymbol{1}_{\{ c < 3\}}, T\boldsymbol{1}_{\{ c = 3\}}\} - t_u^{(0)} + \kappa \Bigr)^{-\delta} \times \\
    &\Bigg(\Big[\delta \ln \Bigl(\min \{t_u^{(c+1)}\boldsymbol{1}_{\{ c < 3\}},T\boldsymbol{1}_{\{ c = 3\}}\} - t_u^{(0)} + \kappa \Bigr)+1 \Bigr]^2 + 1\Bigg)\\
    &- \Bigl(t_u^{(c)} - t_u^{(0)} + \kappa \Bigr)^{-\delta} \Bigg(\Bigl[\delta \ln \Bigl(t_u^{(c)} - t_u^{(0)} + \kappa \Bigr) + 1\Bigr]^2 + 1\Bigg) 
     \Bigg] 
    \end{split} 
\end{equation*}
\begin{equation*}
    \frac{\partial^2{\ell}}{\partial{\psi_c} \partial{\gamma_{c'}}} =
    \begin{cases}
        -\sum_{r = 1}^n \dfrac{k_r \frac{|U_{i_r}(s_r)|}{\sum\limits_{j \in I(s_r)} |U_j(s_r)|}}{\Bigl(\psi_c + \gamma_c \frac{|U_{i_r}(s_r)|}{\sum\limits_{j \in I(s_r)} |U_j(s_r)|} \Bigr)^2} & \text{if $c = c'$}\\
        0 & \text{if $c \neq c'$}
    \end{cases}
\end{equation*}
\begin{equation*}
    \begin{split}
        &\frac{\partial^2{\ell}}{\partial{\psi_c} \partial{\kappa}} = - \sum\limits_{u \in \mathcal{U}^{(c)}} \sum\limits_{i \in \mathcal{I}}\Bigg[ \Bigl(\min \{t_u^{(c+1)}\boldsymbol{1}_{\{ c < 3\}}, T\boldsymbol{1}_{\{ c = 3\}}, z_i \} - t_u^{(0)} + \kappa \Bigr)^{-(1+\delta)} \\
        &- \Bigl(\max \{t_u^{(c)}, y_i \} - t_u^{(0)} + \kappa \Bigr)^{-(1+\delta)} \Bigg]  \\
    \end{split}
\end{equation*}
\begin{equation*}
    \begin{split}
        &\frac{\partial^2{\ell}}{\partial{\psi_c} \partial{\delta}} = -\sum\limits_{u \in \mathcal{U}^{(c)}} \sum\limits_{i \in \mathcal{I}} \frac{1}{\delta^2} \Bigg[\Bigl(\min \{t_u^{(c+1)}\boldsymbol{1}_{\{ c < 3\}}, T\boldsymbol{1}_{\{ c = 3\}}, z_i \} - t_u^{(0)} + \kappa \Bigr)^{-\delta} \times \\
        &\Bigl[\delta \ln \Bigl(\min \{t_u^{(c+1)}\boldsymbol{1}_{\{ c < 3\}}, T\boldsymbol{1}_{\{ c = 3\}}, z_i \} - t_u^{(0)} + \kappa \Bigr)+1 \Bigr]\\
        &- \Bigl(\max \{t_u^{(c)}, y_i \} - t_u^{(0)} + \kappa \Bigr)^{-\delta} \Big[\delta \ln \Bigl(\max \{t_u^{(c)}, y_i \} - t_u^{(0)} + \kappa \Bigr)  + 1 \Big] \Bigg]  \\
    \end{split}
\end{equation*}
\begin{equation*}
    \begin{split}
        &\frac{\partial^2{\ell}}{\partial{\gamma_c} \partial{\kappa}} = - \sum\limits_{u \in \mathcal{U}^{(c)}} \Bigl(\min \{t_u^{(c+1)}\boldsymbol{1}_{\{ c < 3\}}, T\boldsymbol{1}_{\{ c = 3\}}\} - t_u^{(0)} + \kappa)^{-(1 +\delta)} - (t_u^{(c)} - t_u^{(0)} + \kappa)^{-(1+\delta)} \Bigr)
    \end{split}
\end{equation*}
\begin{equation*}
    \begin{split}
        &\frac{\partial^2{\ell}}{\partial{\gamma_c} \partial{\delta}} =-\sum\limits_{u \in \mathcal{U}^{(c)}} \frac{1}{\delta^2} \Bigg[ \Bigl(\min \{t_u^{(c+1)} \boldsymbol{1}_{\{ c < 3\}}, T\boldsymbol{1}_{\{ c = 3\}}\} - t_u^{(0)} + \kappa \Bigr)^{-\delta} \times \\ &\Big[\delta \ln \Bigl(\min \{t_u^{(c+1)}\boldsymbol{1}_{\{ c < 3\}},T\boldsymbol{1}_{\{ c = 3\}}\} - t_u^{(0)} + \kappa \Bigr)+1 \Big] - \Bigl(t_u^{(c)} - t_u^{(0)} + \kappa \Bigr)^{-\delta} \Bigr[\delta \ln \Bigl(t_u^{(c)} - t_u^{(0)} + \kappa \Bigr) + 1\Bigl] \Bigg]  \\
    \end{split}
\end{equation*}
\begin{equation*}
    \begin{split}
        &\frac{\partial^2{\ell}}{\partial{\kappa} \partial{\delta}} = -\sum_{c=0}^{3} \sum_{r = 1}^n \frac{k_{r, c}}{\Bigl(s_r - t_{u_r}^{(0)} + \kappa \Bigr)} + \sum_{c=0}^{3}\sum\limits_{u \in \mathcal{U}^{(c)}} \sum\limits_{i \in \mathcal{I}} \psi_c \times \\
        &\Bigg[\ln \Bigl(\min \{t_u^{(c+1)}\boldsymbol{1}_{\{ c < 3\}}, T\boldsymbol{1}_{\{ c = 3\}}, z_i \} - t_u^{(0)} + \kappa \Bigr) \times \Bigl(\min \{t_u^{(c+1)}\boldsymbol{1}_{\{ c < 3\}}, T\boldsymbol{1}_{\{ c = 3\}}, z_i \} - t_u^{(0)} + \kappa \Bigr)^{-(1 + \delta)} \\
        &- \ln \Bigl(\max \{t_u^{(c)}, y_i \} - t_u^{(0)} + \kappa \Bigr) \times \Bigl(\max \{t_u^{(c)}, y_i \} - t_u^{(0)} + \kappa \Bigr)^{-(1+\delta)} \Bigg]  \\
        &-\sum_{c=0}^{3}\sum\limits_{u \in \mathcal{U}^{(c)}}\gamma_c \Bigg[\ln \Bigl(\min \{t_u^{(c+1)}\boldsymbol{1}_{\{ c < 3\}},T\boldsymbol{1}_{\{ c = 3\}}\} - t_u^{(0)} + \kappa \Bigr) \times \\
        &\Bigl(\min \{t_u^{(c+1)} \boldsymbol{1}_{\{ c < 3\}}, T\boldsymbol{1}_{\{ c = 3\}}\} - t_u^{(0)} + \kappa \Bigr)^{-(1+\delta)} - \ln \Bigl(t_u^{(c)} - t_u^{(0)} + \kappa \Bigr) \times \Bigl(t_u^{(c)} - t_u^{(0)} + \kappa \Bigr)^{-(1+\delta)}  \Bigg]  \\
    \end{split}
\end{equation*}

\section{Goodness-of-fit Tests}
\label{app: Goodness-of-fit Tests}

To assess the goodness-of-fit of the mutually exciting point processes model for the crowdfunding data, we compute the overall compensator $\Lambda(t)$ for each realized contribution time, by considering all unrealized times between all $(u, i)$ pairs, using Equation \ref{eq: Compensator}. Optimization is necessary, to avoid the complexity of three summations, and repeated operations. We can adopt a bottom up dynamic programming approach to compute $\Lambda(t)$ up to time t,  by noticing that the transformed time $t_r^{*} = \Lambda(t_r)$ can be expressed in terms of all previous contributions. In other words, at realized time $t_r$, $r = 2, \ldots, k$ where $k$ is the total number of contributions, only the history between $t_{r-1}$ and $t_{r}$ needs to be considered, as it can be seen in Equation \ref{eq: Dynamic Programming}.
\begin{equation}
\begin{split}
\label{eq: Dynamic Programming}
&\Lambda(t_r) = \Lambda(t_{r-1}) \\
 &-\sum\limits_{u \in U(t_r)} \sum\limits_{i \in I(t_r)} \sum_{\substack{m: \tau_{ui}^{(m)} \geq t_{r-1} \\ m: \tau_{ui}^{(m)} < t_r}}
\left(\psi_{C_u\left(\tau_{ui}^{(m-1)}\right)} + \gamma_{C_u\left(\tau_{ui}^{(m-1)}\right)} \frac{\left|U_i\left(\tau_{ui}^{(m-1)}\right)\right|}{\sum\limits_{j \in I\left(\tau_{ui}^{(m-1)}\right)} \left|U_j\left(\tau_{ui}^{(m-1)}\right)\right|} \right) \times \\
    &\left(\frac{\left(\tau_{ui}^{(m)} - t_u^{(0)} + \kappa\right)^{-\delta}}{\delta} - \frac{\left(\tau_{ui}^{(m-1)} - t_u^{(0)} + \kappa\right)^{-\delta}}{\delta} \right)  \\
     &- \left(\psi_{C_u\left(\tau_{ui}^{(\eta_{ui})}\right)} +\gamma_{C_u\left(\tau_{ui}^{(\eta_{ui})}\right)} \frac{\left|U_i\left(\tau_{ui}^{(\eta_{ui})}\right)\right|}{\sum\limits_{j \in I\left(\tau_{ui}^{(\eta_{ui})}\right)} \left|U_j\left(\tau_{ui}^{(\eta_{ui})}\right)\right|} \right) \times \left(\frac{\left(t_{r} - t_u^{(0)} + \kappa\right)^{-\delta}}{\delta} - \frac{\left(\tau_{ui}^{(\eta_{ui})} - t_u^{(0)} + \kappa \right)^{-\delta}}{\delta} \right) 
\end{split}
\end{equation}
Simplifying Equation \ref{eq: Dynamic Programming} is similar to deriving the expressions of the derivatives in Appendix \ref{app: Maximum Likelihood Estimation}. For the $\psi_{C_u(\cdot)}$ term that is not multiplied by the item relative popularity, only the summation over the users and items need to be taken into account at each $t_r$. For the $\gamma_{C_u(\cdot)}$ term, it can be noted that for each user, the relative popularity of all items sums to 1, so the summation over the items and the relevant times can be omitted. Considering the potential variations in the number of users and items between two consecutive contribution times $t_{r-1}$ and $t_{r}$ is also important, as items can start or end and users may register. Then, the compensator at time $t_r$, $r = 2, \ldots, k$ reduces to 
\begin{equation}
\begin{split}
\label{eq: Dynamic Programming Simplified}
&\Lambda(t_r) = \Lambda(t_{r-1}) \\
&-\sum\limits_{u \in U(t_r)} \sum\limits_{i \in I(t_r)} \psi_{C_u(t_{r-1})} \times \left(\frac{\left(\min \{t_r, z_i \} - t_u^{(0)} + \kappa\right)^{-\delta}}{\delta} - \frac{\left(\max \{t_u^{(C_u(t_{r-1}))}, t_{r-1}, y_i \}- t_u^{(0)} + \kappa\right)^{-\delta}}{\delta} \right) \\
&-\sum\limits_{u \in U(t_r)} \gamma_{C_u(t_{r-1})}\times \left(\frac{\left(t_{r} - t_u^{(0)} + \kappa\right)^{-\delta}}{\delta} - \frac{\left(\max\{t_{r-1}, t_u^{(C_u(t_{r-1}))}\} - t_u^{(0)} + \kappa\right)^{-\delta}}{\delta} \right) 
\end{split}
\end{equation}
with the special case at $r = 1$ for the first contribution, 
\begin{equation}
\begin{split}
\label{eq: Dynamic Programming first contribution}
\Lambda(t_1) &= -\sum\limits_{u \in U(t_1)} \sum\limits_{i \in I(t_1)} \psi_{C_u(t_u^{(0)})} \times \left(\frac{\left(\min \{t_1, z_i \} - t_u^{(0)} + \kappa\right)^{-\delta}}{\delta} - \frac{\left(\max \{t_u^{(0)}, y_i\}- t_u^{(0)} + \kappa\right)^{-\delta}}{\delta} \right) \\
&-\sum\limits_{u \in U(t_1)} \gamma_{C_u(t_u^{(0)})}\times \left(\frac{\left(t_{1} - t_u^{(0)} + \kappa\right)^{-\delta}}{\delta} - \frac{\Bigl(\kappa\Bigr)^{-\delta}}{\delta} \right) 
\end{split}
\end{equation}

\end{appendix}
\end{document}